%% file: NordarXiv.TEX
\documentclass[11pt,fleqn]{article}

\newcommand{\be}{\beq\label}
\newcommand{\ee}{\eeq}

\input preamble.tex

\begin{document}
\twocolumn[
%\jnumber{4}{2015}

\Title{Regularized Lienard-Wiechert fields in a space with torsion}

\Aunames{Vladimir V. Kassandrov\auth{a,1} and Joseph A. Rizcallah\auth{b,2}}

\Addresses{
\addr a {Institute of Gravitation and Cosmology, Peoples' Friendship
	University of Russia, Moscow, Russia}
\addr b {School of Education, Lebanese University, Beirut, Lebanon}
	}

%\Rec{June 25, 2015}

\Abstract
  {A natural modification of the equations of covariantly-constant vector fields (CCVF) in Weyl geometry leads us to consider a metric compatible geometry possessing conformal curvature and torsion fully determined by its trace. The latter is interpreted as a form of the electromagnetic (EM) 4-potentials and, on a fixed metric background, turns out to be fully determined by the CCVF equations.  When the metric is set Minkowskian, the named equations possess two topologically distinct solutions, with the associated EM fields being asymptotically of the Lienard-Wiechert type and having distributed sources, with a fixed (``elementary'') value of the electric charge. One of the solutions is everywhere regular, whereas the other - singular on a 2-dimensional shell. The speed of propagation of EM fields depends on the local charge density and only asymptotically approaches the speed of light.}

] %%%%%%%%%%%%%%%%%%%%%%%%%%%%%%%%%%%%%%%%%%%%%%%%%%%%%%%%%%%%%%%
\email 1 {vkassan@sci.pfu.edu.ru}
\email 2 {joeriz68@gmail.com}

\section{Introduction}

  It is well-known, that all attempts to construct a successful geometric theory of electromagnetism, beginning with Weyl's unified theory~\cite{weyl19}, have met with failure. In particular, Einstein's dream~\cite{Ein} of describing elementary particles by means of nonsingular solutions of nonlinear differential equations~\footnote{From a contemporary standpoint, Einstein's views involved soliton-like field distributions with finite energies and other physical characteristics.} still awaits its neat realization.

An ambitious version of such a geometric unified field theory set out to obtain two {\em distinct} types of regular solutions (with different masses), corresponding to one and the same in magnitude ``elementary" electric charge which could be interpreted as the ``positive and negative electron"~\cite[p. 191]{Viz}, to use the terminology of the time. According to Einstein, it is the {\em overdetermined structure} of the fields' differential equations that could give rise to the discreteness of the particles' physical characteristics~\cite{Ein1} (see also~\cite[p. 197]{Viz}), primarily the ``quantum" of charge. Nothing of the sort has been achieved, neither in Einstein's time (see, e.g.~\cite{pauli21, Viz}) nor at present.

However, forgoing the canonical Lagrangian approach and alternatively seeking field equations within purely geometric structures, in particular among {\em covariantly-constant vector fields}~\cite[p. 553] {ksmh}, may provide new insights and bring us one step closer toward the realization of the aforementioned geometric program.
   
A covariantly constant vector field (CCVF) $K_\mu$, in a space with affine connection $\Gamma^\rho_{\nu\mu}$, is one that remains invariant under the parallel transport effected by $\Gamma^\rho_{\nu\mu}$ in any direction. In other words, the covariant derivative of $K_\mu$ with respect to $\Gamma^\rho_{\nu\mu}$ must vanish:
\be{ccvf}
\nabla_\nu K_\mu : = \partial_\nu K_\mu - \Gamma^\rho_{\nu\mu} K_\rho = 0.
\ee
A CCVF is closely related to the structure of the space's holonomy groups~\cite{Hall,Hall2}. In Riemann geometries, every CCVF is necessarily a Killing vector field, corresponding to some symmetry of the metric $g_{\mu \nu}$, which otherwise is left largely unconstrained.

By contrast, in Weyl geometry~\cite{weyl19,rosen82}, when (the conformal class of) the metric is fixed, the existence of a CCVF imposes stringent restrictions on the nonmetricity vector $A_\mu$ identified traditionally with the electromagnetic (EM) potentials. In this case, the system of CCVF equations (and its fundamental solution) possess an array of remarkable properties~\cite{KasJos14}. In particular, in addition to the classical invariance w.r.t. the conformal-gauge transformation
\beq    \label{wconform}
	g_{\mu\nu} \mapsto \e^{\lambda} g_{\mu\nu},\ \
	A_\mu \mapsto A_\mu -\frac{1}{2}\d_\mu \lambda,\ \
	K_\mu \mapsto K_\mu,
\eeq
inherited from the Weyl geometric structure itself ($\lambda=\lambda(x)$ being an arbitrary differentiable function of coordinates), the system (\ref{ccvf}) in Weyl space exhibits an additional gauge invariance, which resembles gauge symmetry in quantum mechanics, in the sense that only the CCVF and the non-metricity vector are affected while the metric remains intact. 

Moreover, system (\ref{ccvf}) in a Minkowski metric background admits a fundamental solution for which the congruence of null CCVF reproduces the null shear-free congruence of rays generated by  a point-like source on an  arbitrary worldline. The associated non-metricity vector $A_\mu$ gives then rise to the well-known Lienard-Wiechert (LW) solution in classical electrodynamics, with the electric charge being naturally fixed in value to $\pm 1$. As it turns out~\cite{KasJos14}, for the fundamental solution the above mentioned additional symmetry is closely related to the reparametrization group of the point charge's worldline. On the other hand, the charge ``self-quantization" can be traced to the universality of the speed of light, with its sign being associated with the retarded/advanced LW anzatz, hence with the direction of the ``arrow of time", in close analogy with Feynman's well-known representation.    

Nevertheless, despite its mathematical elegance, it is well known that Weyl's unified theory, in its classical form, encounters major interpretation difficulties, related primarily to the change of scale upon parallel transport ~\cite{pauli21,eddington21,Dirac}. Since the origin of most of these difficulties can be traced back to the geometry's nonmetricity, in the present paper we attempt a CCVF in a different kind of geometry -- the so-called Norden geometry~\cite{Norden} -- which is metric compatible, possesses conformal curvature and allows a natural EM interpretation through the trace of the torsion tensor~\cite{Maluf,Kasan}. 

The rest of the paper is organized as follows: in section 2 we motivate and introduce Norden geometry through a natural modification of the system of CCVF in Weyl space. In section 3 we explore some generic properties of the solutions to the CCVF equations in Norden space. Section 4 is devoted to the fundamental solution of the system of CCVF in Norden space and the associated regularized EM fields of the Lienard-Wiechert type. We conclude the paper in section 5 with a brief discussion of the obtained results.

\section{Norden geometry}

  Because of the space's nonmetricity, for any vector field $K_\mu$ which is covariantly constant in Weyl's space, both its contravariant counterpart $K^\mu = g^{\mu\nu}K_\nu$ and its square $K^2:=K_\mu K^\mu$ are not. Note that the latter transforms according to $K^2 \mapsto \exp{(-\lambda)} K^2$ under (\ref{wconform}). To amend this 'asymmetry', we shall try to give up the conformal-gauge invariance of $K_\mu$ in favor of the invariance of its square $K^2$ (compare with~\cite{rosen82}, where the length-invariance has been ensured, more generically, by defining of new parallel transport). This automatically leads us to consider the following transformations (instead of (\ref{wconform})):      
\bearr    \label{nconform}
	g_{\mu\nu} \mapsto \e^{\lambda} g_{\mu\nu},\ \ \
	A_\mu \mapsto A_\mu -\frac{1}{2}\d_\mu \lambda,
\nnn
	K_\mu \mapsto \e^{\lambda/2}K _\mu,\ \ \
	K^\mu \mapsto \e^{-\lambda/2} K^\mu,
\ear
and consequently entails the modification of the CCVF equations, which should now be form-invariant under the new transformations (\ref{nconform}). It is straightforward to check that the equations
\be{nccvf}
\partial_\nu K_\mu = \widehat{\Gamma}^\rho_{\nu\mu}K_\rho - A_\nu K_\mu,
\ee
and
\[
	\d_\nu K^\mu = -\widehat{\Gamma}^\mu_{\nu\rho}K^\rho + A_\nu K^\mu
\]
with $\widehat{\Gamma}^\rho_{\nu\mu} = \gamma^\rho_{\nu\mu}+A_\nu \delta^\rho_\mu+ A_\mu \delta^\rho_\nu-A^\rho g_{\nu\mu}$ being the Weyl connection (and $\gamma^\rho_{\nu\mu}$ being the Cristoffel symbols), are indeed invariant under the new transformations (\ref{nconform}). The above  equations define in fact a new kind of CCVF with both its covariant $K_\mu$ and contravariant $K^\mu$ components being covariantly constant with respect to one and the same connection 
\be{nconn}
\Gamma^{\rho}_{\nu\mu}=\widehat{\Gamma}^{\rho}_{\nu\mu} -A_\nu\delta^{\rho}_{\mu}=\gamma^\rho_{\nu\mu}+A_\mu \delta^\rho_\nu- A^\rho g_{\nu\mu},
\ee
which is no longer symmetric (in the lower indexes) but, unlike Weyl's connection, is compatible with the metric:
\be{met}
\nabla_\rho g_{\mu\nu} = 0,
\ee
where $\nabla$ stands for the covariant derivative with respect to (\ref{nconn}). In the mathematical literature, connection (\ref{nconn}) is said to define a \emph{semi-symmetric} affine geometry~\cite[p. 148]{Norden}.

We have thus naturally arrived at a CCVF set up in an effective Riemann-Cartan geometry with connection (\ref{nconn}), compatible with the metric, and possessing torsion of the special form  
\be{tor}
T^{\rho}_{[\nu\mu]}=\frac{1}{2}\left(A_\mu\delta^{\rho}_{\nu}-A_\nu\delta^{\rho}_{\mu}\right),
\ee
determined solely by its trace  
\be{ttor}
T^{\rho}_{[\nu\rho]}=-\frac{3}{2}A_\nu.
\ee
In view of $A_\mu$'s transformational properties (\ref{nconform}), the latter may be naturally identified, up to a dimensional factor, with the components of the 4-vector of potentials of an effective EM field  $F_{\mu \nu}=\partial_\mu A_\nu-\partial_\nu A_\mu =\nabla_\mu A_\nu-\nabla_\nu A_\mu $. 

Moreover, using $\Gamma^{\rho}_{\nu\mu}=\widehat{\Gamma}^{\rho}_{\nu\mu} -A_\nu\delta^{\rho}_{\mu}$ and the conformal-gauge invariance of the Weyl connection $\widehat{\Gamma}^{\rho}_{\nu\mu}$ , one can easily see that, under transformations (\ref{nconform}), connection (\ref{nconn}) transforms in a gradient fashion:
\be{ncontrans}
\Gamma^{\rho}_{\nu\mu} \mapsto \Gamma^{\rho}_{\nu\mu}+\frac{1}{2}\delta^\rho_\mu\partial_\nu \lambda.
\ee
Notwithstanding, the curvature corresponding to (\ref{nconn}) turns out to be conformal. This becomes manifest from the structure of the curvature tensor $R^\mu_{~\nu \alpha \beta} = \widehat{R}^\mu_{~\nu \alpha \beta}-\delta^\mu_\nu F_{\alpha\beta}$, which differs from Weyl's (conformal) curvature tensor $\widehat{R}^\mu_{~\nu \alpha \beta}$ by the evidently conformal-gauge invariant $\delta^\mu_\nu F_{\alpha\beta}$. The invariance of curvature under (\ref{nconform}) was first established in~\cite{Obukhov}, using fiber bundle considerations, while an EM interpretation of the underlying geometry was offered some time later~\cite{Maluf} within a Lagrangian framework. We therefore see that a unified field theory based on Norden's geometry can reproduce many remarkable features of Weyl's celebrated theory, while at the same time does not suffer from the notorious shortcomings of the latter.

\section{Covariantly constant vector\\ fields in Norden space}

  We now turn to the properties of the solutions of (\ref{nccvf}). First of all, it is worth mentioning that these equations are regarded as a self-consistent system of PDEs from which both the CCVF $K_\mu$ and the 4-vector potential $A_\mu $ are to be determined, while, at this stage, the metric $g_{\mu\nu}$ is assumed to be given (up to an arbitrary conformal factor). Thus, the CCVF system comprises 16 equations for $4+4=8$ unknown functions, and so is overdetermined. As we shall see later, this, among others, leads to ``quantized'' sources of the EM fields associated with the solutions of (\ref{nccvf}). 

Writing (\ref{nccvf}) out and rearranging terms we find 
\be{newccvf}
K_{\mu;\nu} = A_\mu K_\nu - (A^\rho K_\rho)g_{\mu \nu},
\ee
where the semicolon denotes the Riemann covariant derivative w.r.t. the Cristoffel symbols. Contracting (\ref{newccvf}) with $K^\mu$ we readily see that $K^\mu K_{\mu;\nu}=0$, so that every solution $K_\mu$ of (\ref{newccvf}) necessarily has a constant norm, which may or may not be null and which, in view of transformations (\ref{nconform}), provides a global invariant characteristic of the solution $K_\mu$. 
Contracting (\ref{newccvf}) with $K^\nu$ we find $$K_{\mu;\nu}K^\nu = (K^\nu K_\nu)A_\mu -(A^\nu K_\nu)K_\mu,$$ so that every {\it null} CCVF ($K^\nu K_\nu=0$) in Norden's space corresponds to a null geodesic congruence in the underlying Riemannian space. Moreover, taking the antisymmetric part of (\ref{newccvf}) we get $K_{[\mu;\nu]} = A_{[\mu} K_{\nu]}$, which implies that $K_{\mu}=\lambda \partial_\mu \psi$ for some scalars $\lambda$ and $\psi$~\cite[p. 68]{ksmh}. Therefore, any solution $K_\mu$ of (\ref{newccvf}) defines a non-rotating congruence, which is also geodesic, in the sense of the underlying Riemannian geometry, if and only if it is null.

As to the EM potential $A_\mu$ and the associated  EM field strength tensor $F_{\mu \nu}$, the integrability conditions of (\ref{newccvf}) give 
\be{intcondfull}
R^\mu_{~\nu \alpha \beta}K_\mu=0,
\ee
%where $R^\mu_{~\nu \alpha \beta}$ is the curvature tensor,
with $R_{\mu \nu \alpha \beta}:=g_{\mu \tau}R^\tau_{~\nu \alpha \beta}$ being the sum of the Riemann-Christoffel curvature tensor $\bar{R}_{\mu \nu \alpha \beta}$ plus a non-Riemann part $\breve{R}_{\mu \nu \alpha \beta}$ of the form 
\bearr
	\breve{R}_{\mu \nu \alpha \beta} = A_{\mu;\alpha}g_{\nu\beta}
	- A_{\mu;\beta}g_{\nu\alpha} +A_{\nu;\beta}g_{\mu\alpha}
\nnn \ \
	- A_{\nu;\alpha}g_{\mu\beta} + A_\mu A_\beta g_{\nu\alpha}
	- A_\mu A_\alpha g_{\nu\beta}
\nnn \ \
	+ A_\nu A_\alpha g_{\mu\beta} - A_\nu A_\beta g_{\mu\alpha}
\nnn \ \
	+ A^\lambda A_\lambda(g_{\nu\alpha}
	g_{\mu\beta} -g_{\mu\alpha}g_{\nu\beta}),
\earn
which is also antisymmetric in the first and last pairs of indexes. Separating the
fully antisymmetric part of (\ref{intcondfull}), by contracting it with the fully antisymmetric tensor $\epsilon^{\tau\nu\alpha\beta}$ and noting the cyclic relation $\bar{R}_{\mu \nu \alpha \beta} + \bar{R}_{\mu \alpha \beta \nu} + \bar{R}_{\mu \beta \nu \alpha}=0$, we find 
 
\be{intcond}
\tilde{F}^{\mu \nu}K_\nu=0,
\ee
where $\tilde{F}_{\mu \nu}$ is the 2-form dual to $F_{\mu \nu}$ corresponding to a solution $A_\mu$ of (\ref{newccvf}). From (\ref{intcond}) it follows that 
\be{Fsimp}
F_{\mu\nu}=\omega_{[\mu} K_{\nu]},
\ee
for some $\omega_\mu$, and consequently
\be{inv} 
\tilde{F}^{\mu \nu}F_{\mu \nu}=0.
\ee
In the special case of a Minkowskian background with metric $g_{\mu\nu}(x) =\eta_{\mu\nu}$ this implies
\be{EMconst}
\vec{H}=\vec{n} \times \vec{E} \Rightarrow \vec{n} \cdot \vec{H}=0, \vec{E}\cdot \vec{H}=0,	
\ee
where $\vec{E}$ and $\vec{H}$ are the effective electric and magnetic field strengths and $n^a=K^a/K^0, a=1,2,3$, assuming $K^0 \neq 0$.

Maxwell's vacuum equations, which we consider to be $F^{\mu\nu}_{~~;\nu}=0$ (rather than $\nabla_\nu F^{\mu\nu}=0$, which are not conformal-gauge invariant), do not hold for every solution of (\ref{newccvf}). Nevertheless, we can adopt a different perspective and consider the first and second pairs of Maxwell's equations as the definitions of the EM-fields through potentials and a conserved current-density 4-vector, respectively. From this standpoint, Maxwell's equations with the corresponding source will hold identically for every solution of the CCVF system. This approach to the dynamics of EM-fields seems quite legitimate, provided the CCVF system (\ref{newccvf}) admits fundamental solutions of the Coulomb type, i.e. solutions for which the current-density 4-vector asymptotically vanishes and the fields exhibit the correct multipole structure. This will be established in the next section.

\section{The fundamental solution}

\begin{figure}
%\label{fig:hyp}
\centering
\includegraphics[scale=0.6]{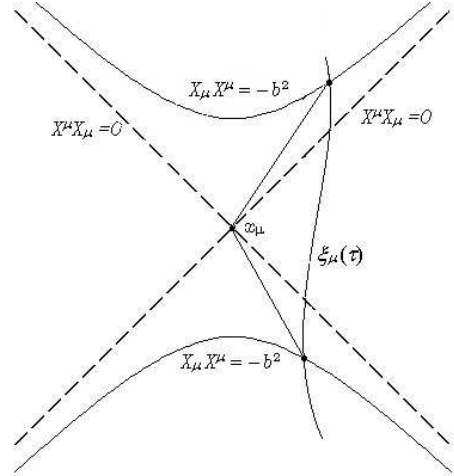}
\caption{The two-sheeted hyperboloid defining $\tau(x)$}\label{fig:hyp}
\end{figure}

  Here, we shall derive a class of solutions of the CCVF system using a construction generalizing the one used to obtain the Lienard-Wiechert fields in classical electrodynamics. To this end, consider an arbitrary worldline $\xi(\tau)$ in ({\em non-flat}, due to a nonzero torsion) Norden space with a Minkowskian background metric $g_{\mu\nu} = \eta_{\mu\nu} = diag\{-1,1,1,1\}$. For a space-time point $x_\mu$ the two-sheeted hyperboloid/light cone equation   
\be{retard}
X_\mu X^\mu = \pm b^2, ~~~~X^\mu: = x^\mu -\xi^\mu(\tau),
\ee
where $b$ is a real nonnegative constant, implicitly defines the field $\tau=\tau(x)$ (compare with~\cite{Bonnor}, see Fig.~\ref{fig:hyp}). 
% --------------------------------------- fig 1
% ---------------------------------------

Differentiating (\ref{retard}) w.r.t. $x_\mu$, one obtains
\be{prtau}
\partial_\mu \tau = \frac{X_\mu}{P},
\ee
provided the quantity $P:=X_\lambda \xi^{\prime\lambda}$ (the analog of the so called retarded distance~\cite[p. 121]{Bonnor}) does not vanish, where the prime~$\prime$~denotes differentiation with respect to $\tau$. Let us now use the fields $\tau(x)$ and $P(x)$ as generating functions for the 
CCVF. Since $K_\mu$ is necessarily nonrotating, one may try $K_\mu=P\partial_\mu \tau = X_\mu$. 
Using (\ref{retard}), one finds then for the derivatives of $K_\mu$   
\be{secderiv}
\partial_\nu K_\mu=\partial_\nu X_\mu= \eta_{\mu\nu} - \xi_\mu^\prime \frac{X_\nu}{P}.
\ee
Now the CCVF equations (\ref{newccvf}) 
should define the potentials $A_\mu(x)$. To simplify calculations, let us first satisfy the contracted equation $\partial_\mu K^\mu = \partial_\mu X^\mu=3$. Contracting the CCVF equations and comparing, we arrive at the relation 
\be{contr}
A^\mu X_\mu =-1,
\ee
so that for the potentials $A_\mu(x)$ one may assume the following Lorentz 
covariant ansatz:
\be{ansatz}
A_\mu = -\frac{\xi_\mu^\prime}{P} =  -\frac{\xi_\mu^\prime}{X_\lambda{\xi^\prime}^\lambda}.
\ee
Together with $K_\mu=X_\mu$ this identically satisfies the full CCVF equations (\ref{newccvf}), as one can readily verify. Unlike the potentials $A_\mu$, the components $K_\mu$ of the CCVF are evidently defined up to a multiplicative constant.

It is noteworthy that equations (\ref{retard}), (\ref{prtau}) and (\ref{ansatz}) are {\em reparametrization invariant}, though, unlike the situation in Weyl geometry~\cite{KasJos14}, the 4-potentials (\ref{ansatz}) do not transform gradient-wise under the transformation $\tau \mapsto \sigma(\tau)$, so the reparametrization group is not explicitly related to the electromagnetic gauge group.

Moreover, the retarded ($X^0 = x^0 - \xi^0 >0$) and advanced ($X^0 = x^0 - \xi^0 <0$) potentials correspond to opposite unit charges in (\ref{ansatz}). A similar result obtains in Weyl's geometry~\cite{KasJos14}. This common feature of both geometries seems to follow directly from the structure of the CCVF equations and automatically calls to mind Feynman's well-known particle/antiparticle representation.
 
For $b=0$, on account of (\ref{retard}), (\ref{ansatz}) is identical to the familiar Lienard-Wiechert potential, generated by a point source moving along the worldline $\xi(\tau)$, with the principal difference that the corresponding dimensionless electric charge is equal to $\pm 1$. In particular, for a source at rest the solution reduces to the Coulomb solution with a unit charge. 

On the other hand, for $b \neq 0$ (\ref{ansatz}) does not generally satisfy Maxwell's vacuum equations. As mentioned earlier, in this case we regard the second pair of Maxwell's equations as the definition of the current and charge densities. In the stationary case, the simplest spherically symmetric solution can easily be found~\cite{Kasan}
\bearr		\label{torsol}
	K_0=\epsilon\sqrt{r^2 \mp b^2}, \ \ \ K_a=x^a;
\nnn
	A_0=\frac{\epsilon}{\sqrt{r^2 \mp  b^2}},\ \ \
	A_a=0,\ \ \ \epsilon = \pm 1.
\ear

% ------------------------------------------ fig 2
\begin{figure}
\centering
\includegraphics[scale=0.35]{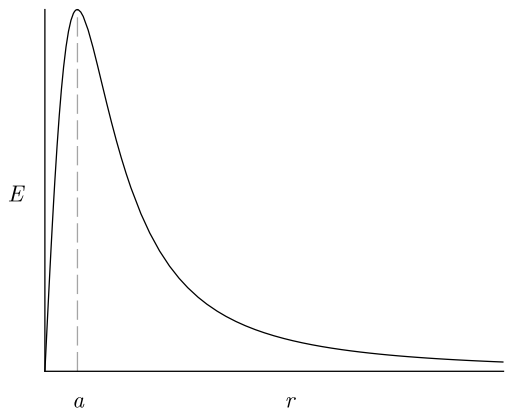}
\caption{The electric field $E$ distribution}\label{fig:sol}
\end{figure}
% ------------------------------------------ fig 3
\begin{figure}
\centering
\includegraphics[scale=0.35]{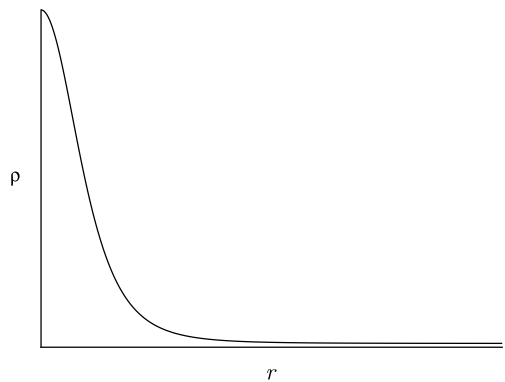}
\caption{The electric charge density $\rho$ distribution}\label{fig:den}
\end{figure}
% ------------------------------------------

We thus obtain \emph{two topologically different types of particle-like solutions}, describing the electrostatic field of a \emph{distributed} charged source. One solution is soliton-like, with no singularities. The corresponding electric field strength $E$ and charge density $\rho$ are depicted in Fig.~\ref{fig:sol},~\ref{fig:den}. Note that $E$ attains a maximum 
at $r=a=b/\sqrt{2}$, which can be regarded as the effective radius of the source. The other solution is singular on the spherical shell $r =b$. %, separating the interior and the exterior.  
As $r \rightarrow \infty$ both solutions exhibit the correct Coulomb asymptotic 
\begin{equation}
\label{coul-asm}
A_0\approx \frac{\epsilon}{r},
\end{equation}
and, as with $b=0$, their electric charge is fixed to unity. 

\begin{figure}
\includegraphics[scale=0.35]{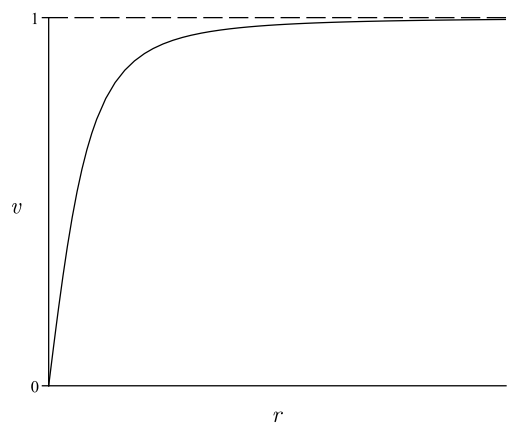}
\caption{The velocity $v$ of EM signals vs. $r$}\label{fig:vr}
\end{figure}

In the case of the regular solution, a rather unusual situation arises. If one assumes that the center of charge is located at the origin $r=0$, then the time it takes a signal to propagate from the origin to the field point $\vec{r}$ turns out to be a nonlinear function of $r$, namely $\sqrt{r^2+b^2}$, bounded from below by $b$. It thus follows that the speed of propagation $v$ is a function of $r$ given by (see Fig.~\ref{fig:vr}) 
% ------------------------------------------ fig 4
% ------------------------------------------
$$v=r/\sqrt{r^2+b^2}$$ 
and ranges between $0$, for very close to the center of charge field points $r \rightarrow 0$, and the ``speed of light'' $1$, for very far ones $r \rightarrow \infty$.
We can interpret this fact as the presence of an effective {\it field-like medium}, which on the one hand, gives rise to the source, and on the other hinders the propagation of EM signals. 

\section{Conclusion}

  In the paper we present an alternative to Weyl's theory of electromagnetism, arising naturally from the modification of the CCVF equations in Weyl's geometry. Unlike the latter, the obtained conformal geometry - Norden's geometry - is defined by a {\em metric connection} and possesses a torsion tensor fully determined by its trace, which we adopt as the basic object for an EM interpretation of the theory.

The congruence associated with the CCVF, characterized by the invariant and constant norm $K_\mu K^\mu$, turns out to be nonrotating and in the null case ($b=0$) also geodesic in the sense of the underlying Riemannian geometry. Although Maxwell equations in free space don't follow from the considered CCVF system, the integrability conditions of the latter restrict the associated 2-forms to those for which the invariant $\tilde{F}^{\mu \nu}F_{\mu \nu}$ vanishes identically.

%used in special relativity to derive the Lienard-Wiechert potential for an arbitrary moving point charge, 

A straightforward modification of the well-known light cone method based on the retardation equation is devised to arrive at the fundamental solution of the CCVF system in the special case of a (conformally) Minkowskian metric. The EM structure associated with the solution is identified as a {\em deformed} Lienard-Wiechert field of an arbitrarily moving charge. Using this result, it is shown that the CCVF system in Norden space offers various descriptions of the particles' structure, i.e. sources of EM-fields. 
%It is argued, that for these solutions, the retarded and advanced potentials exhibit a close connection to the sign of the electric charge.
For $b=0$ the Coulomb structure of the fundamental solution dictates the classical picture of a point-like source, whereas the nonnull case ($b \neq 0$) provides two topologically different types of {\em distributed} sources: a soliton-like ansatz and a solution singular on a shell, both exhibiting the correct Coulomb asymptotic. An unexpected feature of the soliton-like solution is the dependence of the speed of propagation of EM signals on the properties of the {\em effective medium}, particularly its charge density.

Although it is still premature to discuss the physical significance of the obtained solutions, the latter possess at least some of the remarkable features anticipated by Einstein (see, e.g.~\cite{Ein, Viz}), most notably, the nonsigular EM fields associated with arbitrarily moving distributed sources and the emergence, from an overdetermined system of field equations, of two asymmetric solutions with a Lienard-Wiechert (Coulomb, in the static case) asymptotic and ``elementary'' value of the electric charge. We have also shown that the sign of the electric charge is associated with the ``arrow of time'', thus naturally including the {\em particle/antiparticle asymmetry} in the theory.   

Finally, it is interesting that the difference in the topology of solutions (\ref{torsol}) may give rise to very different masses of the corresponding particles. This issue, however, presupposes a complete resolution of the general interaction problem. In particular, it is necessary to augment the CCVF equations to a full system, from which it would be also possible to determine the metric (which we have so far assumed to be Minkowskian). Some of the  problems arising in the search of such a {\em self-consistent} approach (including agreement with classical GR tests) are discussed in~\cite{KasJos14}.

\section*{Acknowledgements}
The authors are grateful to Yu. N. Obukhov for pointing out his interesting paper~\cite{Obukhov} and other literature dealing with the properties of the above considered geometry.

\end{document}

%% file: preamble.tex
\usepackage{amsfonts,amssymb,cite}
\usepackage{epsf,graphicx}

\def\noi{\noindent}

\newcommand{\Title}[1]{\noi {{\Large\bf #1}}\\[1ex]}

\def\Aunames#1{\noi{\bf #1}}
\def\auth#1{${}^{#1}$}
\def\Addresses#1{\medskip\noi \protect
	\begin{description}\itemsep -3pt {\it #1} \end{description}}
\def\addr#1#2{\item[${}^{#1}$]{\it #2}}

\newcommand{\Abstract}[1]{\vskip 2mm \begin{center}
        \parbox{16.4cm}{\small\noi #1} \end{center}\medskip}

\def\email#1#2{\footnotetext[#1]{e-mail: #2}\addtocounter{footnote}{1}}

\def\nqq{\hspace*{-2em}}

\def\Jl#1#2{#1 {\bf #2},\ }

\def\ApJ#1 {\Jl{Astroph. J.}{#1}}
\def\CQG#1 {\Jl{Class. Quantum Grav.}{#1}}
\def\DAN#1 {\Jl{Dokl. AN SSSR}{#1}}
\def\GC#1 {\Jl{Grav. Cosmol.}{#1}}
\def\GRG#1 {\Jl{Gen. Rel. Grav.}{#1}}
\def\JETF#1 {\Jl{Zh. Eksp. Teor. Fiz.}{#1}}
\def\JETP#1 {\Jl{Sov. Phys. JETP}{#1}}
\def\JHEP#1 {\Jl{JHEP}{#1}}
\def\JMP#1 {\Jl{J. Math. Phys.}{#1}}
\def\NPB#1 {\Jl{Nucl. Phys. B}{#1}}
\def\NP#1 {\Jl{Nucl. Phys.}{#1}}
\def\PLA#1 {\Jl{Phys. Lett. A}{#1}}
\def\PLB#1 {\Jl{Phys. Lett. B}{#1}}
\def\PRD#1 {\Jl{Phys. Rev. D}{#1}}
\def\PRL#1 {\Jl{Phys. Rev. Lett.}{#1}}

\def\lal{&&\nqq {}}

\def\beq{\begin{equation}}
\def\eeq{\end{equation}}
\def\bear{\begin{eqnarray}}
\def\bearr{\begin{eqnarray} \lal}
\def\ear{\end{eqnarray}}
\def\earn{\nonumber \end{eqnarray}}

\def\nnn{\nonumber\\ \lal }

\def\e{{\,\rm e}}
\def\d{\partial}